\input{aipcheck.tex}

  \def\selectedoptions{final}
\documentclass[
   \selectedoptions
  ]
  {aipproc}

\layoutstyle{8x11double}

\SetInternalRegister\hbadness{8000} % pseudo latin isn't breaking very well :-)

\newcommand\doingARLO[2][]{%
  \ifx\mmref\undefined #1\else #2\fi
}

\begin{document}

\title 
      [ ]
      {Gamma-ray bursts: the tip of the iceberg?}

\classification{43.35.Ei, 78.60.Mq}
\keywords{Document processing, Class file writing, \LaTeXe{}}

\author{Maurice H.P.M. van Putten}{
  address={MIT 2-378, 77 Massachusetts Avenue, Cambridge, MA 02139},
  email={mvp@schauder.mit.edu},
  thanks={This work was commissioned by the AIP}
}

% \copyrightholder{Acoustical Scociety of America}
\copyrightyear  {2001}

\begin{abstract}
The spin-energy $E_{rot}$ of a Kerr black hole surrounded by a torus may 
power emissions in multiple windows. The recently determined true GRB-energy
$E_\gamma=3-5\times 10^{50}$erg indicates a minor fraction $E_j/E_{rot}\simeq0.1\%$ in
baryon poor output, here 
considered as jets along open magnetic flux-tubes from the horizon to infinity.
A major fraction $E_{gw}/E_{rot}\simeq5\%$ is expected in gravitational radiation from the
torus. A LIGO/VIRGO detection of $\alpha=2\pi \int fdE_{gw}$ in excess of the 
neutron star limit $\alpha^*\simeq0.005$ promises a calorimetric test for Kerr black holes.
We expect a sample of LIGO/VIRGO detections to obey the distribution of redshift corrected 
GRB-durations.
\end{abstract}

\date{\today}

\maketitle

\section{Introduction}

%\ifthenelse{\equal\selectedlayoutstyle{6x9}}{\par\bfseries 
%  Note: The entire paper will be reduced 15\% in the printing
%  process. Please make sure all figures as well as the text within the
%  figures are large enough in the manuscript to be readable in the
%  finished book.\par\bfseries 
%  Note: The entire paper will be reduced 15\% in the printing
%  process. Please make sure all figures as well as the text within the
%  figures are large enough in the manuscript to be readable in the
%  finished book.\par\bfseries 
%  Note: The entire paper will be reduced 15\% in the printing
%  process. Please make sure all figures as well as the text within the
%  figures are large enough in the manuscript to be readable in the
%  finished book.\normalfont}{}

Black hole-torus systems may represent high-energy astrophysical transient sources.
They feature the prospect of multi-window emissions powered by the spin energy 
$E_{rot}$ of a Kerr black hole.
This could take the form of outflows along the axis 
of rotation accompanied by emissions from the torus in various channels:
gravitational radiation, Poynting flux-dominated and baryonic winds and, when 
sufficiently hot, neutrino emissions.
Ultimately, these systems may provide definitive tests for 
Kerr black holes as objects in Nature -- the most compact
energy reservoirs in angular momentum.

Rotating black holes where discovered by Kerr as exact solutions to general 
relativity \cite{kerr63}.
The specific angular momentum of their radiation is at least twice that of
the black hole, which suggests that Kerr black holes may be luminous under
appropriate conditions. Identifying Kerr black holes
will require observational evidence for its defining properties (see \cite{mvp01f}):
{a compact horizon surface} in common with non-rotating Schwarzschild black 
holes; {frame-dragging of space-time} is described by an angular velocity $-\beta$
of zero angular momentum observers;
{a compact energy reservoir} of energy $E_{rot}=2M\sin^2(\lambda/4)$, where
$a/M=\sin\lambda$ denotes the ratio of specific angular momentum $a$ to the 
black hole mass $M$. In an extreme Kerr black hole, about half of the rotational energy 
corresponds to the top ten percent of the angular velocity $\Omega_H=\tan(\lambda/2)/2M$.

Black hole-torus systems harboring Kerr black holes are leading candidates as the inner 
engine of gamma-ray bursts (see \cite{mvp01a} for a review). 
GRBs are characteristically non-thermal in the 
range of a few hundred keV with a bi-modal distribution in durations of 
short bursts around 0.3s and long bursts around 30s \cite{kou93}. 
Black hole plus disk or torus systems may represent failed-supernovae 
\cite{woo93} hypernovae \cite{pac98} or 
black hole-neutron star coalescence \cite{pac91}, where the former is 
intimately connected to star forming regions \cite{pac98,blo00}. 
With GRBs remnants potentially found in Soft X-ray transients \cite{bro00}
GRO J1655-40 \cite{isr99} and V4641Sgr \cite{oro01}, the putative black hole
assumes the observed mass-range $3-14M_\odot$ (Fig. 1). 
We recently identified long/short bursts with rapidly/slowly spinning black holes
in a state of suspended-/hyperaccretion \cite{mvp01b}. A mean de-redshifted
duration on the order of tens of seconds
corresponds to the life-time of rapid spin in suspended
accretion in the presence of superstrong magnetic fields.

\begin{figure}
\caption{Shown is the distribution of black hole masses in X-ray novae.
	 The top four are XTE J118+408 \cite{mcc01}
	 V4641 Sgr \cite{oro01}, 4U 1543-47 \cite{oro98}
	 and Nova Vel 1993 \cite{fil99}; the lower seven
	 are from  \cite{bai98}. This mass distribution 
	 manisfests a certain diversity in black hole masses of about
	 $3-14M_\odot$. [Reprinted from van Putten, {\em Physics Reports},
	 345 \copyright 2001 Elsevier B.V.]
}
\includegraphics[height=.395\textheight]{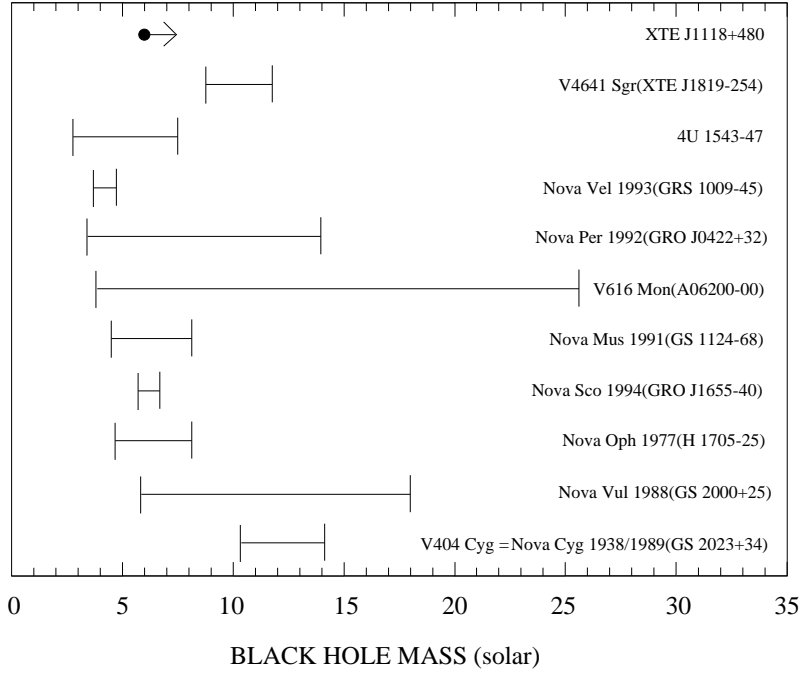}
\end{figure}

Here, we focus on baryon poor jets along the axis of rotation 
along with gravitational radiation from the torus. 
As proposed input to GRBs, the former will represent a minor fraction
of $E_{rot}$ as inferred from the recently determined true GRB-energies
\cite{fra01,pir01}. The latter is expected to be a major fraction of the output
which could be representative for $E_{rot}$.
We thus expect LIGO/VIRGO to detect a distribution of durations in gravitational waves 
which corresponds to the presently observed redshift-corrected distribution of 
GRB-durations.

\section{Multi-window emissions}

A Kerr black hole is expected to be luminous over {\em all} horizon angles, in 
response to a generally uniform magnetic flux. Thus, we expect emissions along
its axis of rotation as well as into the surrounding matter. A detailed analysis
is based on the topology of the magnetic field, partly
by equivalence to pulsar magnetospheres and the
formation of open flux-tubes. 

In its lowest energy state, a rotating black hole surrounded by a torus magnetosphere
develops an equilibrium magnetic moment \cite{mvp01a}
\begin{eqnarray}
\mu_H\simeq aBJ_H,
\label{EQN_MU}
\end{eqnarray}
where $B$ denotes the mean of the poloidal magnetic field in the surrounding 
torus magnetosphere. It corresponds to the Wald equilibrium charge \cite{wal74}
consistent with the no-hair theorem \cite{car68,mvp01a}. It carries over to a largely
force-free magnetosphere around a black hole \cite{lee01} and, in scaling, is analogous
to the equilibrium charge on a neutron star \cite{coh75}.
The black hole hereby maintains essentially a maximal and uniform
horizon flux at arbitrary rotation rates. 

\subsection{Baryon-poor jets}
A rapidly rotating black hole may support an open flux-tube supported by 
Eqn.(\ref{EQN_MU}).
These are endowed with slip/slip- and
ingoing/outgoing-boundary conditions on the horizon/infinity.
The charge-density about the axis of rotation of the black hole
satisfies $\rho=-(\Omega+\beta)B/2\pi$ (see \cite{mvp01a} for references), 
where $\Omega$ denotes the angular velocity
of the open flux-tube; a sign-change from positive in a lower section attached to 
the black hole to negative in the semi-infinite section above occurs at some
height above the black hole when $0<\Omega<\Omega_H$. This permits a continuous
current along the open flux-tube from infinity into the hole, with outflow to
infinity and inflow into the black hole.
Open magnetic flux-tubes are a remarkable natural phenomenon, perhaps
most dramatically demonstrated by solar activity.
Magnetic mediated outflows in general, therefore, require the creation of open 
magnetic flux-tubes which formally extend to infinity. 
Open magnetic flux-tubes may be created from a torus magnetosphere around a black hole, 
by change in topology \cite{mvp01a}. 
This change in topology represents transient fast magnetosonic wave, 
which might be excited as a nonlinear feature to strong Alfv\'en waves or by
superradiance within the torus magnetosphere. It produces a coaxial 
structure of two flux-tubes, an inner tube supported by the equilibrium magnetic moment 
of the black hole and an outer tube supported by the torus. The flux in the
inner/outer tube is $\pm2\pi A_\phi$ in terms of the vector potential $A_\phi$. 
The outer flux-tube endowed with no-slip/slip boundary conditions on the
surface of the torus/infinity.

%\newpage
The inner flux-tube forms an powerful artery for the spin-energy of the black hole.
In asymptotically charge-separation equilibrium, it assumes an angular
velocity $\Omega_+$ on the horizon and $\Omega_-$ at larger distances by the
slip-slip boundary conditions. These lower and upper sections are separated by 
differential rotation. For a net flux $2\pi A_\phi$, 
the ingoing boundary conditions produce a current $I_+=(\Omega_H-\Omega_+)A_\phi$ 
in the small angle approximation \cite{pun90,mvp01a}; 
likewise, the current at infinity is $I_-=\Omega_-A_\phi$ for
ultrarelativistic winds \cite{mvp01a}. This leaves a Faraday-induced potential 
$V=(\Omega_+-\Omega_-)A_\phi$ along the differentially rotating inner tube.
Global current closure is over the outer flux-tube, which corotates
with the angular velocity $\Omega_T$ of the torus by its no-slip boundary conditions.
By current continuity, differential rotation hereby creates a baryon-poor jet with
luminosity \cite{mvp01a}
\begin{eqnarray}
L_j=\frac{1}{2}\Omega_T(\Omega_H-2\Omega_T)A_\phi^2.
\label{EQN_LP}
\end{eqnarray}

\begin{figure}
\caption{
Shown is the distribution of redshift corrected durations, obtained from 10 GRBs with
individually determined redshifts from their afterglow emissions 
(GRB000926,GRB000418,GRB000301c,GRB990510,GRB990123,GRB980613,
GRB980425,GRB971214,GRB970508,GRB970228). This distribution represents the
life-time of the inner engine. In the black hole-torus model, the proposed 
gravitational radiation from the torus is simultaneous with the baryon-poor output 
powering the GRB. LIGO/VIRGO detections of these emissions (from cosmologically
nearby sources) are expected to obey a similar distribution. 
	 }
\includegraphics[height=.395\textheight]{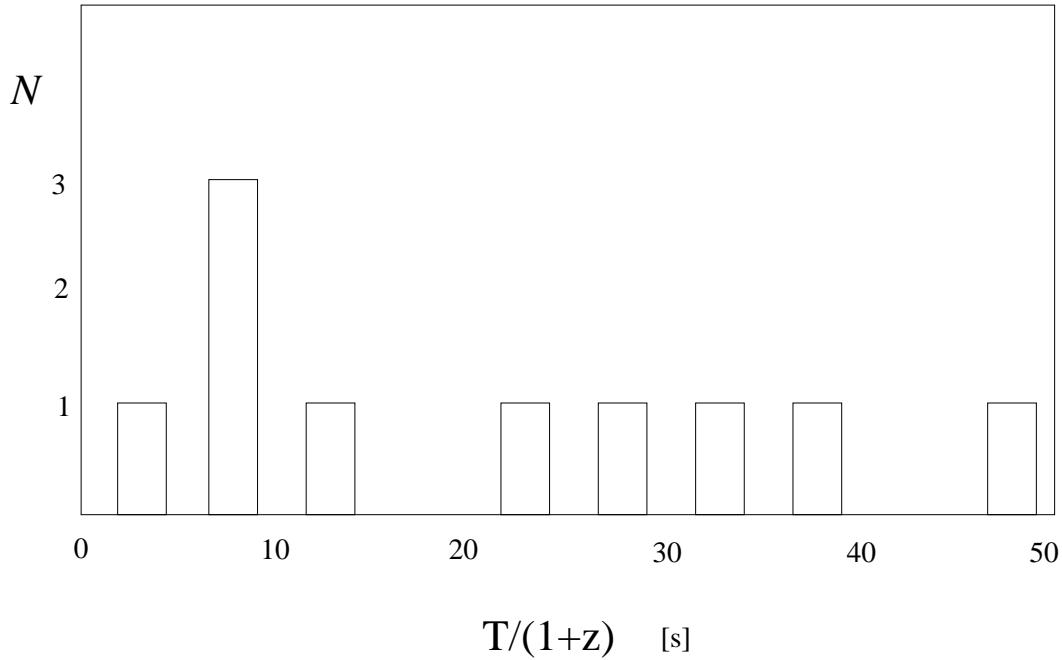}
\end{figure}

These low-$\sigma$ outflows may represent the baryon-poor input
to cosmological gamma-ray bursts (GRBs).
When the low-$\sigma$ outflow is hidden from view by an
intervening medium, the observed spectrum will be different from black body
radiation. The observed spectrum - but less so the characteristic energies -- 
depends on the subsequent evolution of this medium and its interaction 
with the interstellar medium. In particular, the non-thermal GRB emissions may 
hereby be the testimonial to a hypernova progenior wind, associated with its
rapid rotation and consistent with the recently discovered iron line emissions
\cite{piro01}. This introduces shocks and synchrotron radiation.
The observed emissions, therefore, should be non-thermal with approximately
similar characteristic energies as emitted from the gap. The detailed
structure of the emissions further depends on the
interaction of this accelerated wind mass with the interstellar medium 
\cite{pir98}.

The recenlty inferred true GBR-energies of $3-5\times 10^{50}$erg \cite{fra01,pir01}
imply $E_j/E_{rot}\simeq0.01\%/\epsilon$ in baryon poor outflows.
Here $\epsilon\simeq0.15\%$ denotes the radiation efficiency, assuming 
$E_{rot}\simeq M/3$ with $M=7M_\odot$. As a specific application of (\ref{EQN_LP}), 
this indicates an open flux-tube with half-opening angle $\theta_H\simeq 35^o$
\cite{mvp01c}. The
half-opening angles on the celestial sphere will satisfy $\theta_j\le \theta_H.$
The duration of the burst and the observed half-opening
angles $\theta_j$ may be uncorrelated, upon collimation by external parameters such as 
hydrostatic static pressure in a GBR-precursor winds or the interstellar medium, or
positively correlated, upon collimation by winds \cite{lev00} coming off the torus.

\subsection{Gravitational radiation}

A surrounding torus receives energy and angular momentum by equivalence in
poloidal topology to pulsar magnetospheres: the inner face of a torus around a 
black hole receives energy and angular momentum, as does a pulsar when infinity wraps
around it. The outer face looses angular momentum and energy to infinity, as does a pulsar 
in flat space-time. These equivalences becomes apparent when working in a frame of references
fixed to the horizon of the black hole and, respectively, infinity (Mach's principle).
When the black hole spins rapidly, it develops a state of suspended accretion for the duration of 
rapid spin of the black hole.  The high incidence of the 
black hole-luminosity onto the inner face indicates that the emissions from the torus may be
luminous. We thus find an output 
\begin{eqnarray}
E_{gw}=1-2\%M
\end{eqnarray}
in gravitational radiation --
about two orders of magnitude higher than the inferred true GRB-energies.

The major fraction $E_{gw}/E_{rot}\simeq 5\%$ 
emitted at twice the Keplerian frequency, i.e., $f=1-2$kHz, promises black hole-torus systems
to be viable sources for the upcoming broadband gravitational wave observatories LIGO \cite{abr92} and
VIRGO \cite{bra92}. Thus, black hole-torus systems may have a compactness parameter \cite{mvp01f}
\begin{eqnarray}
\alpha=2\pi \int_0^{E_{gw}} fdE> \alpha^*\simeq 0.005
\end{eqnarray}
in excess of the limit for rapidly spinning neutron stars.
This provides for the first time a {\em calorimetric} compactness test for Kerr black holes.
The proposed association to GRBs predicts that a future sample of LIGO/VIRGO detections will
satisfy a distribution of durations which obeys the distribution of redshift corrected 
GRB-durations $T/(1+z)$ (Fig. 2).
The displayed spread in $T/(1+z)$ is consistent with the narrow mass range of $3-14M_\odot$
in SXTs. Indeed, we expect a positive correlation between $T/(1+z)\propto M^2$
and $E_{gw}\propto M$ (as well as $E_{j}$ and $E_\gamma)$ in view of
$E_{rot}\propto M$ and a black hole-to-torus coupling $\propto M^{-1}$ for a 
universal ratio of poloidal magnetic field energy-to-kinetic energy in the torus \cite{cow01}.

\section{Summary}

We have described a prospect for multi-window emissions from Kerr black holes powered by
its rotational energy. Surrounded by a torus, a Kerr black hole is luminous over {\em all} horizon angles 
in its lowest energy state. These systems are long-lived in a state of suspended accretion, which 
operates by equivalence in poloidal topology to pulsar magnetospheres. Quite generally, the powerful 
competing torques acting on the torus introduce turbulent shear flow, which may stimulate the 
formation of a quadrupole moment in its mass distribution.
We expect a minor fraction in baryon-poor jets from a differentially rotating tube along the axis of 
rotation and a major fraction in gravitational radiation from the torus. (Further output is expected
in torus winds and, when sufficiently hot, neutrino emissions.)
These black hole-torus systems are predicted to be powerful LIGO/VIRGO-sources of gravitational radiation, 
permitting for the first time a calorimetric compactness test for Kerr black holes. A sample of
LIGO/VIRGO detections is predicted to obey the distribution of redshift corrected GRB-durations.

\begin{theacknowledgments}
This work is partially supported by NASA Grant 5-7012, and MIT C.E. Reed Award and a NATO 
Collaborative Linkage Grant. The author thanks E. Costa and A. Levinson for stimulating 
discussions.
\end{theacknowledgments}

%\newpage
%\centerline{Figure Captions}
%\mbox{}\\
% choose bibtex style depending on layout style and options used in
% sample:
\doingARLO[\bibliographystyle{aipproc}]
          {\ifthenelse{\equal{\AIPcitestyleselect}{num}}
             {\bibliographystyle{arlonum}}
             {\bibliographystyle{arlobib}}
          }
\bibliography{sample}
\end{document}